\documentclass[twocolumn, nofootinbib]{revtex4}

\usepackage{amsmath,amsfonts,amssymb}
\usepackage{mathrsfs}
\usepackage{graphicx}
\usepackage[english]{babel} 
\usepackage{url} 
\usepackage{color}
\usepackage{bbold}
\usepackage{adjustbox}
\usepackage[utf8]{inputenc} 
\usepackage{float}
\usepackage[colorlinks=true,citecolor=blue,urlcolor=blue]{hyperref}

\usepackage{amsmath}
\usepackage{wasysym}
\usepackage{amssymb}
\usepackage{amsthm}
\usepackage{physics}
\usepackage{booktabs}
\usepackage[dvips,letterpaper,margin=1in,bottom=0.7in]{geometry}
\usepackage{tensor}
\usepackage{mathtools}
\usepackage{mathrsfs}

\usepackage{enumerate}
\renewcommand{\v}[1]{\mathbf{#1}} 
\usepackage{bm}
\newcommand{\gv}[1]{\bm{#1}}
 
\let\baraccent=\= 
\renewcommand{\=}[1]{\stackrel{#1}{=}} 

\usepackage{cancel}

\theoremstyle{definition}

\usepackage{esint}
\usepackage{booktabs}
\usepackage{enumitem}
\usepackage{graphicx}
\usepackage{longtable}
\usepackage{float}
\usepackage{chemfig}
\graphicspath{ {/home/john/Homework/Spring 2022/OChemLab/Reports/Chemlum/Images/} }
\numberwithin{equation}{section}

\begin{document}

\title{Multi-axionlike description of the dark sector in light of the Hubble and \(\sigma_8\) tensions}

\author{John Dumancic\({}^{1}\)}
\email{dumancjp@mail.uc.edu}
\affiliation{\({}^{1}\)Department of Physics, University of Cincinnati, Cincinnati, OH 45221 USA\\
\({}^{2}\)Department of Physics and Astronomy, Northwestern University, Evanston, IL 60208 USA}
\author{Richard Gass\({}^{1}\)}
\email{gassrg@ucmail.uc.edu}
\affiliation{\({}^{1}\)Department of Physics, University of Cincinnati, Cincinnati, OH 45221 USA\\
\({}^{2}\)Department of Physics and Astronomy, Northwestern University, Evanston, IL 60208 USA}
\author{Ennis Mawas\({}^{1,2}\)}
\email{ennismawas2030@u.northwestern.edu}
\affiliation{\({}^{1}\)Department of Physics, University of Cincinnati, Cincinnati, OH 45221 USA\\
\({}^{2}\)Department of Physics and Astronomy, Northwestern University, Evanston, IL 60208 USA}
\author{L.C.R. Wijewardhana\({}^{1}\)}
\email{rohana.wijewardhana@gmail.com}
\affiliation{\({}^{1}\)Department of Physics, University of Cincinnati, Cincinnati, OH 45221 USA\\
\({}^{2}\)Department of Physics and Astronomy, Northwestern University, Evanston, IL 60208 USA}

\begin{abstract}
    Local methods of direct determination of the Hubble constant and \(\sigma_8\) seem to conflict with the predictions made from the cosmic microwave background and \(\Lambda\)CDM. We propose a proof-of-concept model that models portions of the dark sector as several coupled axion-like fields, resulting in both early and late time departures from \(\Lambda\)CDM. We determine that the model successfully eliminates both the Hubble and \(\sigma_8\) tensions, while remaining consistent with both the DESI survey and the BAO sound horizon.
\end{abstract}

\maketitle

\section{Introduction} \label{sec:i}
\noindent The $\Lambda$CDM framework is the prevailing model for describing the composition of the universe, including dark energy, matter, and radiation. It provides a robust explanation for a variety of observations, such as the cosmic microwave background (CMB), the accelerated expansion of the universe, and predictions consistent with general relativity \cite{Condon:2018eqx}. Despite its successes, recent observational and theoretical challenges to $\Lambda$CDM have prompted the investigation of alternative models. A key issue is the Hubble tension, a statistically significant $4\sigma$ to $6\sigma$ discrepancy between measurements of the Hubble parameter $H_0$ derived from early-time (CMB) data \cite{2020,Dutcher:2021vtw,DAmico:2019fhj,Alam:2020sor,Zhang:2018air,Pogosian:2020ded} and late-time (local) observations \cite{Riess:2024vfa,Riess_2011,Riess_2016,Riess_2019,Riess:2020fzl, Breuval:2020trd,Camarena:2019moy,Soltis:2020gpl,Huang:2019yhh,Pesce:2020xfe,freedman2024statusreportchicagocarnegiehubble,Li:2024pjo}. This persistent conflict has inspired numerous extensions to the standard $\Lambda$CDM model, including early- and late-time modifications to dark energy \cite{Lin:2019qug,Hill:2020osr,Sakstein:2019fmf,Niedermann:2020dwg, Aghanim:2018eyx,DiValentino:2019dzu,Yang:2020zuk,Cai:2021wgv,Wang:2016och}, and interacting dark-energy and dark-matter models \cite{DiValentino:2017iww,Yang:2018uae,DiValentino:2019ffd,Yang:2018euj,Yang:2019uzo,Wang:2016lxa}. One of the persistent challenges in modern cosmology is reconciling the sound horizon constraints from baryon acoustic oscillations (BAO) and Type Ia supernova data (Pantheon) with the local measurements of the Hubble constant $H_0$ \cite{DiValentino:2021izs,Knox:2019rjx,Arendse:2019hev}. Late-time solutions often struggle to simultaneously address these tensions, while early-time solutions, although capable of reducing the sound horizon, can introduce conflicts with local observations. To bridge this gap, we propose an axion-based model that has the potential to effectively address the $H_0$ and $\sigma_8$ discrepancies, while remaining consistent with the aforementioned observations.

\noindent Axions, originally introduced as a solution to the strong CP problem in quantum chromodynamics \cite{Peccei:1977hh,Weinberg:1977ma,Wilczek:1977pj,Dine:1981rt,Kim:1979if,Shifman:1979if,Zhitnitsky:1980tq,Marsh:2015xka}, have emerged as compelling candidates for dark matter \cite{PRESKILL1983127,ABBOTT1983133,DINE1983137,Arias:2012az,Visinelli:2017imh} and dark energy \cite{Hlozek:2014lca,Visinelli:2018utg}. Their versatility makes them particularly well-suited for models aimed at modifying the universe's expansion history. Axion-like particles (ALPs), in particular, are a natural extension of this framework \cite{Marsh:2010wq}, with intriguing cosmological applications due to their unique coupling properties and the potential to influence dynamics on various timescales \cite{mawas2021interactingdarkenergyaxions, DAmico:2016jbm}.

\noindent In this study, we examine a model that incorporates both early- and late-time evolving dark-energy axions (without cosmological constant) to modify the expansion history of the universe before and after matter-radiation decoupling. This allows us to increase $H_0$, without violating BAO measurements of the sound horizon.
The model we investigate is based on an interacting-ALP framework \cite{mawas2021interactingdarkenergyaxions, DAmico:2016jbm}. Here, we extend the ALP model proposed by \cite{DAmico:2016jbm} to incorporate an ALP with a potential of the form $V(\phi) = [1-\cos(\phi/f)]^n$ \cite{Kaloper:2016}. 
We present a specific choice of parameters that alleviates both the tension of \(H_0\) and the tension of \(\sigma_8\), a discrepancy between CMB and large-scale structure (LSS) measurements of the linear power spectrum amplitude, $\sigma_8$, on the scales $8 h^{-1} \text{Mpc}$ \cite{Macaulay:2013swa,Ade:2015xua,2020,Heymans:2020gsg,Abbott:2017wau}. Using a method from Barros et al. \cite{Barros:2018efl} to approximate $\sigma_8$ in coupled quintessence models, we will show that our model alleviates the discrepancy in $\sigma_8$ .

\noindent This paper is organized as follows: in Sec. \ref{sec:ii}), we outline the general formalism of $N$ interacting ALPs and their evolution throughout cosmic history, including the evolution of the Hubble parameter. In Sect. \ref{sec:iii}, we discuss matter perturbations in this model and assess the potential to address the $\sigma_8$ tension. In Sect. (\ref{sec:iv}), we identify parameters for the two- and three-axion models that alleviate the Hubble tension. In Sect. \ref{sec:iv.A}, we discuss the methods used to analyze the background evolution and present our results. In Sect. (\ref{sec:iv.B}), we discuss the sound-horizon problem and the limitations of the model. Finally, we present our conclusions in Sect. \ref{sec:v}.

\section{Background Formalism} \label{sec:ii}
\noindent Following \cite{mawas2021interactingdarkenergyaxions}, we consider a model with an arbitrary number of scalar fields minimially coupled to gravity: 
\begin{align}
    L=\sqrt{-g}\left[\frac{1}{2}R-\frac{1}{2}\delta^{ij}g^{\mu\nu}\partial_\mu\phi_i\partial_\nu\phi_j-V(\gv{\phi})\right]\label{eq:ii.1}
\end{align}
From Eq. \ref{eq:ii.1}, we obtain the general equations of motion:
\begin{align}
    R_{\mu\nu}-\frac{1}{2}Rg_{\mu\nu}&=\delta^{ij}\partial_\mu\phi_i\partial_\nu\phi_j-g_{\mu\nu}V(\gv{\phi})\notag\\
    &-\frac{1}{2}g_{\mu\nu}\delta^{ij} g^{\alpha\beta}\partial_\alpha\phi_i\partial_\beta\phi_j\label{eq:ii.2}\\
    &:= T_{\mu\nu}\notag\\
    0&=g^{\mu\nu}\partial_\mu\partial_\nu\phi_i+\pdv{V(\gv{\phi})}{\phi_i}\label{eq:ii.3}
\end{align}
Following  \cite{mawas2021interactingdarkenergyaxions} and \cite{Kaloper_2016}, we define the potential as 
\begin{align}
V(\gv{\phi})&=\sum_{i=1}^N \mu_i^4\left[1-\cos\left(\frac{\phi_i}{f_i}\right)\right]^{n_i}\notag\\
&+\sum_{i=1}^{N-1}\sum_{j=i+1}^N\mu^4_{i,j}\left[1-\cos\left(\frac{\phi_i}{f_i}-c_{i,j}\frac{\phi_j}{f_j}\right)\right]\label{eq:ii.4},
\end{align}
where the \(\phi_i\) is the \(i\)th real scalar field, \(f_i\) is the \(i\)th decay constants, and \(\mu_i\) is the \(i\)th self-interaction constant. The interactions between the \(i\)th and \(j\)th fields are described by the interaction strength constant \(\mu_{i,j}\) and the mixing constant \(c_{i,j}\), which is taken to be an integer to maintain the shift symmetry \(\phi_i\to\phi_i+2\pi\). Motivated by \cite{mustafaAmin}, we have introduced \(n_i\) to generalize the normal axion potential (which ours reduce to when \(n_i=1\)). 
\begin{figure}
    \centering
    \includegraphics[width=\linewidth]{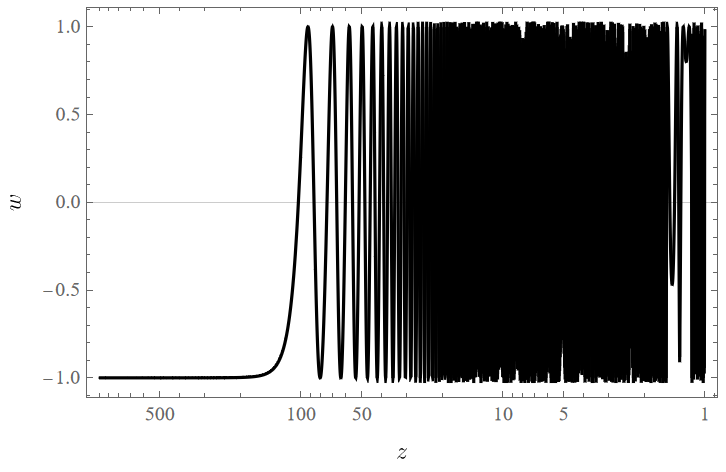}
    \includegraphics[width=\linewidth]{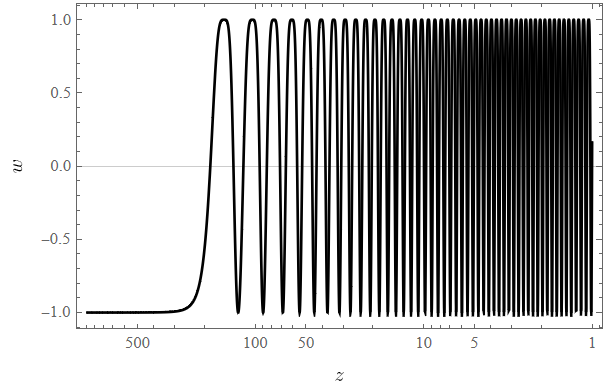}
    \includegraphics[width=\linewidth]{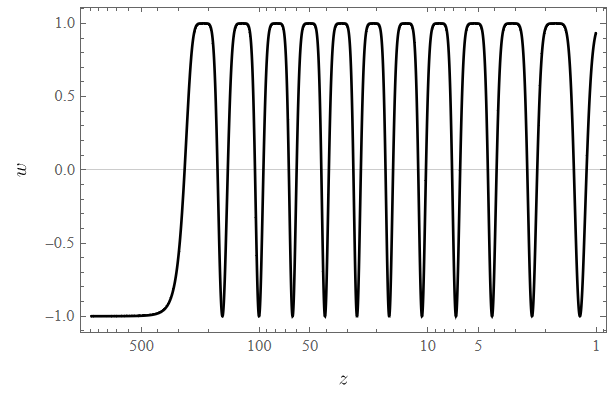}
    \caption{\(w\) for a generalized axion (in this case, \(n=1,2,3\), with \(\ev{w}=0,\tfrac{1}{3},\tfrac{1}{2}\) respectively); for sufficiently steep potential wells the average kinetic dominates the average potential energy, resulting in a higher average \(w\).}
    \label{fig:ws}
\end{figure}
Assuming an Friedmann–Lemaître–Robertson–Walker (FLRW) spacetime, the equations of motion reduce to the Raychaudhuri and Klein-Gordon equations:
\begin{align}
    \dot{H}&=-\frac{1}{2}(\rho+p)\label{eq:ii.6}\\
    0&=\ddot{\phi}_i+3H\dot{\phi}_i+\pdv{V(\gv{\phi})}{\phi_i},\label{eq:ii.7}
\end{align}
where the Hubble parameter obeys the constraint \(H^2=\rho/3\).
Within this framework, the contributions of ALPs to the energy density and pressure of the universe can be characterized as follows:
\begin{align}
    \rho_\phi&=\frac{1}{2}\dot{\gv{\phi}}^2+V(\gv{\phi})\label{eq:ii.8}\\
    p_\phi &=\frac{1}{2}\dot{\gv{\phi}}^2-V(\gv{\phi})\label{eq:ii.9}.
\end{align}

These expressions highlight how ALPs influence cosmic evolution by contributing dynamically to both energy density and pressure.  Ordinary axions typically transition from behaving like dark energy to behaving as dark matter as the universe evolves. Generalized axions, however, exhibit a more diverse set of behaviors depending on the form of their potential. For instance, these ALPs initially act as dark energy but later transition into a fluid with an equation of state parameter \(w_i\) that depends on the steepness of their potential. As shown in \cite{PhysRevD.28.1243,Scherrer:2008be,Norton:2020ert,Poulin:2018cxd}, this behavior can be characterized by:
\begin{equation} 
\ev{w_i}=\frac{n_i-1}{n_i+1}\label{eq:ii.5}. 
\end{equation}
This may be intuitively derived as follows: long after the transition of a given ALP \(\phi\), the \(3H\dot{\phi}\) term in Eq. (\ref{eq:ii.7}) is negligible, and the system behaves effectively as a one-dimensional oscillator where the energy is divided between kinetic and potential contributions.
For a scalar field oscillating in a potential \(V(\phi) = A\phi^{2n}\), the kinetic energy is given by \(T = \frac{1}{2}\dot{\phi}^2\). The total energy of the system is therefore:
\begin{equation}
    E = \frac{1}{2}\dot{\phi}^2 + A\phi^{2n}.
\end{equation}
The energy of the system \(E\) is conserved within our approximation. Averaging over one oscillation period, the total energy can be expressed as:
\begin{equation}
    E = \ev{T} + \ev{V}.
\end{equation}
Averaging \(w\) over a period, we have 
\begin{equation}
\ev{w}=\ev{\frac{T-V}{E}}=E^{-1}\left(\ev{T}-\ev{V}\right).
\end{equation}
According to the virial theorem, we have \(\ev{T}=n\ev{V}\), and thus an equation of state 
\begin{equation}
\ev{w}=\frac{\ev{T}-\ev{V}}{\ev{T}+\ev{V}}=\frac{n-1}{n+1}.
\end{equation}
This result shows that the average equation of state parameter depends solely on the power \(n\) of the potential \(V(\phi) \).\newline
\noindent Returning to our model, it is convenient to write Eqs. (\ref{eq:ii.6}--\ref{eq:ii.7}) in terms of the redshift rather than cosmic time. We have, for some arbitrary function \(f\),
\begin{align}
 \dot{f}&=-H(z)(1+z)f'\label{eq:ii.10}\\
 \ddot{f}&=H(z)(1+z)\left[H'(z)(1+z)f'\right.\label{eq:ii.11}\\
 &+\left.H(z)f'+H(z)(1+z)f''\right],\label{eq:ii.12}
\end{align}
where a prime denotes a derivative with respect to the redshift. 
With this in mind, assuming that there exists in the present day matter and radiation fractions \(\Omega_{m,0},\,\Omega_{r,0}\), we may write down the full energy density and pressure as 
\begin{align}
    \rho(z)&=3H_0^2\left[\Omega_{m,0}(1+z)^3+\Omega_{r,0}(1+z)^4+\frac{\rho_\phi}{3H_0^2}\right]\label{eq:ii.13}\\
    \rho_\phi(z)&=\frac{1}{2}H^2(1+z)^2\delta^{ij}\phi_{i}'\phi_j'+V(\gv{\phi})\label{eq:ii.14}
\end{align}
\begin{align}
    p(z)&=3H_0^2\left[\frac{1}{3}\Omega_{r,0}(1+z)^3+\frac{p_\phi}{3H_0^2}\right]\label{eq:ii.15}\\
    p_\phi(z)&=\frac{1}{2}H^2(1+z)^2\delta^{ij}\phi_{i}'\phi_j'-V(\gv{\phi})\label{eq:ii.16}.
\end{align}
Substituting Eqs. (\ref{eq:ii.13}--\ref{eq:ii.16}) into Eqs. ~(\ref{eq:ii.6}—\ref{eq:ii.7}), we have the following equations of motion:
\begin{align} 
    (1+z)H'&=\frac{1}{2}(\rho+p)\label{eq:ii.17}\\
    0&=H^2(1+z)^2\phi_i''+\pdv{V(\gv{\phi})}{\phi^i}\notag \\
    &+(1+z)H\left[(1+z)H'-2H\right]\phi_i'.\label{eq:ii.18}
\end{align}
For ease of numerical solution, we scale the parameters as follows: 
\begin{equation} 
\begin{aligned}
    \tilde{H}(z)&=H_0^{-1}H(z)\\
    \tilde{f}_i&=M_p^{-1}f_i\\
    \tilde{V}(\gv{\phi})&=(H_0M_p)^{-2}V(\gv{\phi})
\end{aligned}
\;\;\;
\begin{aligned}
    x_i&=f_i^{-1}\phi_i\\
    \tilde{\mu}_i&=(H_0M_p)^{-2}\mu_i\\
    \tilde{\mu}_{i,j}&=(H_0M_p)^{-2}\mu_{i,j}\label{eq:ii.19}
\end{aligned}
\end{equation}
Defining the deacceleration parameter 
\begin{align}
    q:&=\frac{1}{2}\left[1+\sum_{i,j}\frac{1}{2}(1+z)^2\tilde{H}^2\delta^{ij}\tilde{f}_i\tilde{f}_j x_i'x_j'\right.\notag\\
    &+\left.\tilde{H}^{-2}\left(\Omega_{r,0}(1+z)^4-\tilde{V}(\v{x})\right)\right],\label{eq:ii.20}
\end{align}
we may now write Eqs.~\ref{eq:ii.17}—\ref{eq:ii.18} in a numerically tractable form: 
\begin{align}
    (1+z)\tilde{H}'&=(q+1)\tilde{H},\label{eq:ii.21}\\
    -\pdv{\tilde{V}}{x_i}&=\tilde{H}^2\tilde{f}_i(1+z)\left[(1+z)x_i''+(q-1)x_i'\right].\label{eq:ii.22}
\end{align}
\section{Perturbations:} \label{sec:iii}
\noindent To compute \(\sigma_8\), we must consider how matter perturbations evolve within the constraints of this model. From \cite{DAmico:2016jbm}, the evolution equations are as follows:
\begin{align}
0&=\ddot{\delta}_m + 2 H  \dot{\delta}_m- \frac{3}{2} H^2 \Omega_m \delta_m \notag\\
&-\delta^{ij} \left(2\dot{\phi}_i \dot{\delta\phi_j} -  \frac{\partial V}{\partial \phi_i}\delta\phi_i\right)\label{eq:iii.1} \\
0&=\ddot{\delta\phi}_i+ 3 H \dot{\delta\phi}_i + k^2(1+z)^2 \delta \phi_i \notag\\
&+\delta\phi_j \frac{\partial^2 V}{\partial \phi_i \partial \phi_j}- \dot{\delta}_m\dot{\phi}_i,\label{eq:iii.2}
\end{align}
Converting to redshift and scaling the variables, we have
\begin{align}
    0&=\tilde{H}^2(1+z)\left[(1+z)\delta_m''+q\delta_m'\right]-\frac{3}{2}\Omega_{m,0}(1+z)^3\delta_m\notag\\
    &-2\tilde{f}_i^2\tilde{H}^2(1+z)^2x_i'\delta x_i'-\delta x_i \pdv{\tilde{V}}{x_i}\label{eq:iii.3}\\
    0&=\tilde{H}^2(1+z)\left[(1+z)\delta x_i''+(q-1)\delta x_i'\right]\notag\\
    &+\tilde{k}^2(1+z)^2\delta x_i+\frac{\delta x_j}{\tilde{f}_i^2}\pdv[2]{\tilde{V}}{x_i}{x_j}-\tilde{H}^2(1+z)^2\delta_m'x_i'.\label{eq:iii.4}
\end{align}
We define 
\begin{equation}
\delta_m:=\frac{1}{2}\sqrt{k^3P(k)},\,\tilde{k}:=k/H_0,\label{eq:iii.5}
\end{equation}
where \(P(k)\) is the matter power spectrum. 
\section{Results} \label{sec:iv}
\noindent To rectify both the Hubble and \(\sigma_8\) tensions while respecting essential cosmological constraints, we propose a three-field model: two fields are coupled traditional axions with \(n_1=n_2=1\), while the third is an ALP with \(n_3=3\) (\(n_i\) as defined in Eq. (\ref{eq:ii.4}). As seen in Figure (\ref{fig:model}), the two coupled fields essentially behave as time-varying dark energy, while the third field behaves as a fluid with \(w=\tfrac{1}{2}\). Because of this, the mass bounds of \cite{amin2024lowerbounddarkmatter} are not applicable, as no sector of the system behaves as dark matter. Combined, the influence of the three axions successfully resolves both the Hubble and the \(\sigma_8\) tensions, while agreeing with the constraints of BAO and CMB on the acoustic sound horizon. Note that our model is consistent with potential non-\(\Lambda\)CDM evolution of the early universe compatible with current DESI data \cite{2024arXiv240403002D}. 
\subsection{Hubble Tension} \label{sec:iv.A}
\begin{figure}[t]
    \centering
    \includegraphics[width=\linewidth]{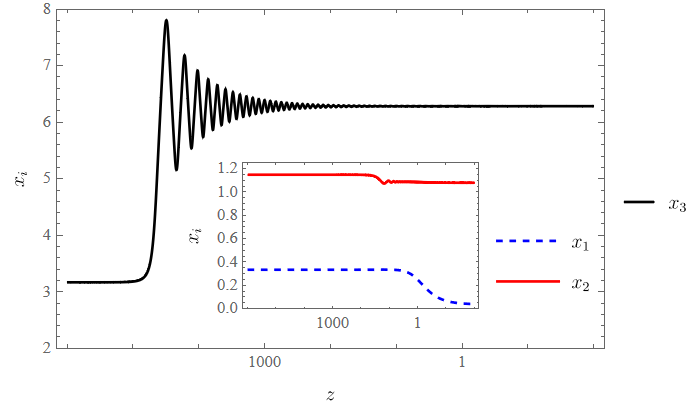}
    \caption{The field values as a function of redshift. \(x_i=\phi_i/f_i\) is the dimensionless axion field. The damping of the amplitude for field \(x_{3}\) is due to the redshift caused by the expanding universe, and it converges to a potential minimum at \(x_3\approx2\pi\).}
    \label{fig:model}
\end{figure}

\begin{figure}[t]
    \centering
    \includegraphics[width=\linewidth]{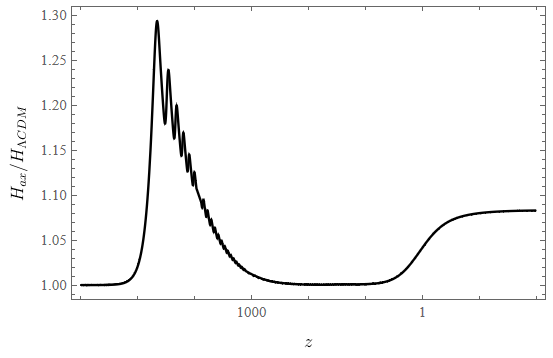}
    \caption{Ratio of the computed Hubble parameter to the Hubble parameter predicted by \(\Lambda\)CDM.}
    \label{fig:Background}
\end{figure}

\noindent To analyze the Hubble tension, we first determine the background evolution. We solve Eqs. (\ref{eq:ii.21}, \ref{eq:ii.22}) as an initial value problem, beginning at \(z_0=10^6\). We take \(\Omega_{m,0}\) and \(\Omega_{r,0}\) from Planck \cite{2020} and specify the initial field values \(x_{i,0}\) and other parameters in Table \ref{tab:ALP_params}. At early times, when \(H\) is large, \(3H\dot{\phi}_i\gg(\partial V/\partial \phi_i)\) in Eq. (\ref{eq:ii.7}), freezing the axion until the two terms are comparable (see Figure (\ref{fig:model})) \cite{DAmico:2016jbm}. So, we may assume that \(\phi_i'(z_0)=x_i'(z_0)=0\). To close the system, we make use of the Friedman equation \(H^2=\rho/3\) \cite{Weinberg:2008zzc}: invoking Eqs. (\ref{eq:ii.13}, \ref{eq:ii.14}) and recalling that \(x_i'(z_0)=0\), we have 
\begin{align}
    \tilde{H}(z_0)&=\bigg[\Omega_{m,0}(1+z_0)^3+\Omega_{r,0}(1+z_0)^4\notag\\
    &+\left.\frac{1}{3}V(\v{x}(z_0))\right]^{1/2}.
\end{align} 
As a function of redshift, the ratio between the dimensionless Hubble constant for \(\Lambda\)CDM and our axion model is shown in Figure (\ref{fig:Background}). The early time deviation from \(\Lambda\)CDM is due to \(\phi_3\), while the late time deviation is due to the coupled \(\phi_1,\phi_2\) system.
\begin{table*}
	\begin{ruledtabular}
		\begin{tabular}{cccccccccccc}
			& $\tilde{\mu}_1^4$	
			& $\tilde{\mu}_2^4$	
			& $\tilde{\mu}_3^4$	
			& $\tilde{\mu}_{1,2}^4$	
			& $c_{1,2}$
			& $\tilde{f}_1$	
			& $\tilde{f}_2$	
			& $\tilde{f}_3$	
			& $x_{1, 0}$	
			& $x_{2, 0}$	
			& $x_{3, 0}$	
			\\ \hline
			& $10$ 
			& $4.81$	
			& $1.5 \times 10^{14}$
			& $30$
			& $35$
			& $0.88$	
			& $0.68$
			& $0.14$	
			& $0.33$	
			& $1.15$
			& $3.17$	
		\end{tabular}
	\end{ruledtabular}
    \caption{\label{tab:ALP_params}Parameters used in numerical solutions to Eqs. (\ref{eq:ii.17}, \ref{eq:ii.18}).  Parameters are scaled per Eq. (\ref{eq:ii.19}). The third ALP has a large energy relative to the first two to effect the early-time modifications.}
\end{table*}

\begin{figure}[t]
    \centering
    \includegraphics[width=\linewidth]{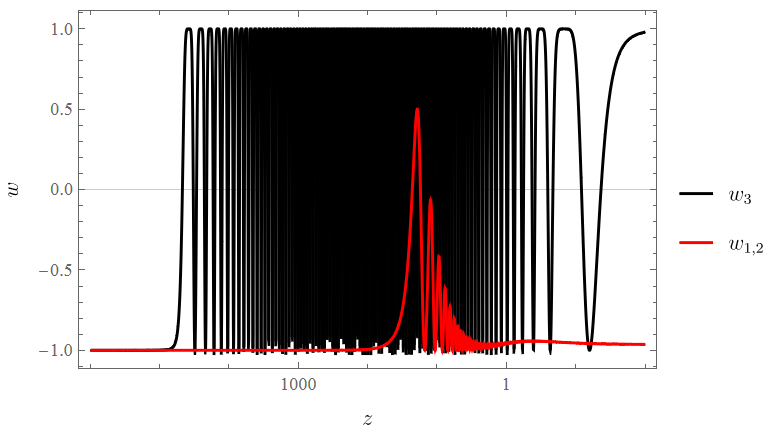}
    \caption{The equations of state for each ALP component of the dark sector of the model. Since \(x_1\) and \(x_2\) are coupled, they are considered as one system. }
    \label{fig:ws}
\end{figure}
\begin{figure}[t]
    \centering
    \includegraphics[width=\linewidth]{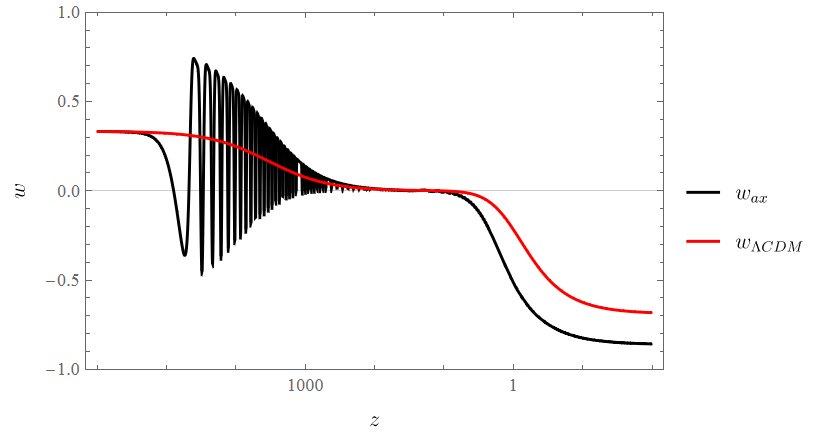}
    \caption{The total equation of state including the matter and radiation fractions, both for the \(\Lambda\)CDM and axion models.}
    \label{fig:ws}
\end{figure}
\noindent We define the sound horizon, as usual \cite{Weinberg:2008zzc}, by 
\begin{align}
    r_d &= \int^\infty_{z_*}\dd{z}\frac{c_s(z)}{H(z)},\label{eq:iv.1}\\
    c_s(z)&:= c\left[3\left(1+\frac{\rho_b}{\rho_\gamma}\right)\right]^{-1/2}.\label{eq:iv.2}
\end{align}
With these, our model gives, subject to the proposed parameters, 
\begin{align}
    H_0&=73.01 \text{ Mpc/(km/s)}\label{eq:iv.3}\\
    r_dh_\text{ax}&=100.06 \text{ Mpc}.\label{eq:iv.4}
\end{align}
These agree to one standard deviation with the experimental measurements of BAO and SH0ES \cite{Zhang:2018air,Pogosian:2020ded}:
\begin{align}
    (H_0)_\text{SH0ES}&=(73.3\pm0.8) \text{ Mpc/(km/s)}\label{eq:iv.5}\\
    (r_dh)_\text{BAO}&=(99.95\pm 1.20) \text{ Mpc/(km/s)}.\label{eq:iv.6}
\end{align}
With the choice of our initial conditions, our Hubble parameter differs from the Planck value by 0.7\% at recombination.
\subsection{\texorpdfstring{\(\sigma_8\text{ }\)} TTension} \label{sec:iv.B}
\noindent Calculating \(\sigma_8\) can be done directly from the matter power spectrum. CAMB \cite{Lewis_2000} is equipped to compute the power spectra of a single ALP added to \(\Lambda\)CDM, with initial conditions from Planck \cite{2020}. The early-time ALP and the two late-time axions are decoupled, each being negligible when the other is active, as shown in Figure (\ref{fig:Background}). This decoupling allows us to use CAMB to compute the power spectrum at \(z=20\) for the ALP + \(\Lambda\)CDM system to an excellent approximation. From this point, we may use this power spectrum as initial conditions for Eqs. (\ref{eq:iii.1} and \ref{eq:iii.2}), which can be numerically solved by Mathematica using the full model; the power spectra can then be extracted using Eq. (\ref{eq:iii.5}).
\begin{figure}[H]
    \centering
    \includegraphics[width=1\linewidth]{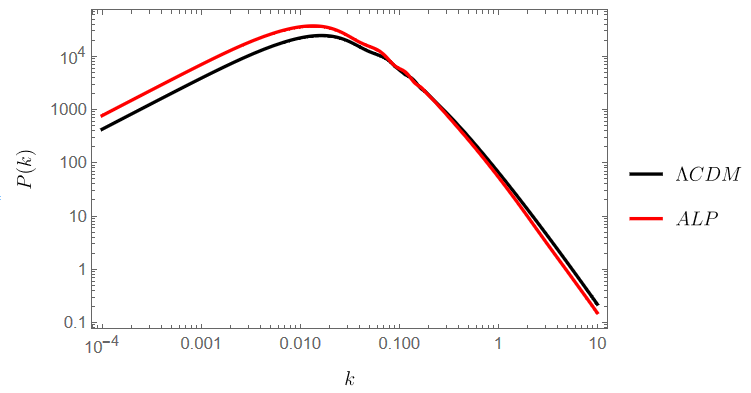}
    \caption{The computed power spectrum along with the unmodified prediction from \(\Lambda\)CDM.}
    \label{fig:power_spectra}
\end{figure} 
\indent\\
\indent\\
\indent\\
\noindent\(\sigma_8\) can then be computed directly from the power spectrum as \cite{Weinberg:2008zzc} \begin{equation}
\sigma_8:=\frac{1}{2\pi^2}\int_0^\infty\dd{k}k^2 P(k) W(8h^{-1},k)^2,\label{eq:iv.7}
\end{equation}
where \(W(r,k)\) is the window function \begin{equation}
W(r,k)=\frac{3}{(kr)^3}\left(\sin kr - kr\cos kr\right).\label{eq:iv.8}
\end{equation}
From this, we determine 
\begin{equation}
\sigma_8=0.812.\label{eq:iv.9}
\end{equation} 
This value cannot be directly compared to experiment, as the cited values by KiDS \cite{Heymans:2020gsg} and DES \cite{Abbott:2021bzy} are predicated on \(\Lambda\)CDM; \(\sigma_8\) is measured at a variety of redshifts between \(z=0\) to \(z=1\) and then propagated to \(z=0\). To find the new value in our model, we propagate the \(\sigma_8\) values measured by KiDS and DES to an average value of \(\bar{z}=0.5\) using \(\Lambda\)CDM and then continue propagating to the present using our model to obtain comparable values. For this propagation, we follow \cite{Barros:2018efl} and utilize the growth function, defined as
\begin{equation}
g(z):=\frac{\delta_m}{\delta_0}.\label{eq:iv.10}
\end{equation}
KiDS gives its results using \(S_8:=\sigma_8\sqrt{\Omega_{m}/0.3}\). Denoting \(\Lambda\)CDM quantities with a `\(\Lambda\)' subscript and quantities from our model with an `ax`' subscript, we have 
\begin{equation}
S_{8,\Lambda}(\bar{z})=S_{8,\Lambda}g_{\Lambda}(\bar{z})\sqrt{\frac{\Omega_{m,\Lambda}(0)}{{\Omega_{m,\Lambda}(\bar{z})}}}.\label{eq:iv.11}
\end{equation}
From this, we obtain 
\begin{equation}
S_{8,\text{ax}}(0)=S_{8,\Lambda}\frac{g_{\Lambda}(0)}{g_\text{ax}(0)}\sqrt{\frac{\Omega_{m,\text{ax}}(0)}{{\Omega_{m,\Lambda}(0)}}}\sqrt{\frac{\Omega_{m,\Lambda}(\bar{z})}{\Omega_{m,\text{ax}}(\bar{z})}}.\label{eq:iv.12}
\end{equation}
Finally, we may obtain \(\sigma_8\) as 
\begin{equation}
\sigma_{8,\text{ax}}=S_{8,\Lambda}\frac{g_{\Lambda}(0)}{g_\text{ax}(0)}\sqrt{\frac{0.3}{{\Omega_{m,\Lambda}(0)}}}\sqrt{\frac{\Omega_{m,\Lambda}(\bar{z})}{\Omega_{m,\text{ax}}(\bar{z})}}.\label{eq:iv.13}
\end{equation}
KiDS and DES give, once transformed,
\begin{align}
    (\sigma_8)_\text{KiDS}&=0.799^{+0.020}_{-0.017}\label{eq:iv.14}\\
    (\sigma_8)_\text{DES}&=0.799^{+0.017}_{-0.017},\label{eq:iv.15}
\end{align}
consistent with our prediction. 
\section{Conclusion} \label{sec:v}
In this study, we proposed a model for the dark sector of the universe, incorporating multiple coupled axion-like fields to address two prominent challenges in cosmology: the Hubble tension and the $\sigma_8$ tension. By leveraging the properties of axion-like potentials, our model effectively modifies the expansion history of the universe at both early and late times. These modifications allow for an increased value of $H_0$, consistent with local observations, while simultaneously alleviating discrepancies in $\sigma_8$ between the cosmic microwave background (CMB) and large-scale structure measurements.

Our results demonstrate that the proposed model is consistent with key cosmological constraints, including baryon acoustic oscillations (BAO) and the sound horizon at recombination, as well as current data from the DESI survey. Furthermore, the introduction of a multi-field framework allows flexibility in addressing multiple tensions without compromising the overall consistency of the model with $\Lambda$CDM at recombination.

This work serves as a proof-of-concept, showcasing the potential for axion-based models to simultaneously resolve both early- and late-time cosmological tensions. Future studies could explore the parameter space more extensively, investigate the implications for the gravitational potential following Amin et al. \cite{Amin_2012}, and consider additional observational signatures that could further test the validity of this approach. 
\section{Acknowledgements} \label{sec:vi}
We would like to thank M. Amin, P. Argyres, C. Bischoff, J. Eby, L. Street, and P. Suranyi for interesting discussions. The research of J.D. and L.C.R.W. is partially supported by the US. Department of Energy grant DE-SC1019775.

\bibliography{dark_energy.bib}

\end{document}